# Review and Recent Advances in PIC Modeling of Relativistic Beams and Plasmas

Brendan B. Godfrey[1, 2, a)]

[1]University of Maryland, College Park, MD 20742, USA
[2] Lawrence Berkeley National Laboratory, Berkeley, California 94720, USA

[a)]Corresponding author: brendan.godfrey@ieee.org

**Abstract.** Particle-in-Cell (PIC) simulation codes have wide applicability to first-principles modeling of multidimensional nonlinear plasma phenomena, including wake-field accelerators. This review addresses both finite difference and pseudo-spectral PIC algorithms, including numerical instability suppression and generalizations of the spectral field solver.

## INTRODUCTION AND SUMMARY

Particle-in-Cell (PIC) simulation codes solve the Vlasov equation by following hundreds of thousands of particles and their associated electromagnetic fields as they evolve in time: Particles are advanced a time-step based on the fields, and then the fields are advanced a time-step based on the currents generated by the particles. This simple process is repeated thousands to millions of times to simulate complex, nonlinear, multidimensional plasma phenomena. Fields are evaluated either on a spatial mesh (FDTD - finite difference time domain) or as a set of spatial Fourier modes (PSTD - pseudo-spectral time domain). Particles, on the other hand, are distributed at arbitrary positions across the mesh, requiring interpolation between fields and particles. PIC codes with explicit temporal algorithms are inherently numerically unstable, and successfully employing them requires increasing numerical instability growth times until they are much longer that the relevant time scales of the physical phenomena simulated. Explicit vacuum field solvers themselves impose Courant limits, $\Delta t < \alpha \Delta x$, with $\Delta x$ the characteristic dimension of a mesh cell and $\alpha$ a numerical factor typically somewhat less than one. PIC methods are discussed in detail in the textbook by Birdsall and Langdon.[1]

Even with the Courant limit satisfied, explicit PIC codes are unstable due to field-particle interpolation coupled with disparities between the Eulerian field solver and Lagrangian particle pusher. Dispersion relations for numerical instabilities can be derived much as they are for physical instabilities and now exist for multi-dimension relativistic beam (or stationary plasma simulated in a relativistically translating frame) simulations. Solutions of numerical dispersion relations indicate that the numerical Cherenkov instability is particularly serious for relativistic beams or plasmas but also suggest various ways to ameliorate it. Several mitigation techniques that involve short wavelength digital filtering combined with minor modifications to the interpolation process have been demonstrated to work well in practice, effectively controlling numerical instabilities in most instances.

This year, generalizations of the PSTD algorithm have been developed that provide variable Courant limits and other flexibilities. The utility of these generalizations currently is being investigated. Some of the techniques developed to ameliorate the numerical Cherenkov instability for the usual PSTD algorithm also work well for these generalized algorithms.

# BRIEF INTRODUCTION TO PIC SIMULATONS

The particle equations of motion in a PIC code can be represented as

$$\frac{d\mathbf{x}}{dt} = \mathbf{v}, \quad \frac{d\gamma\mathbf{v}}{dt} = \sum_{n,l}(\mathbf{E}_l^n + \mathbf{v} \times \mathbf{B}_l^n)W(\mathbf{x} - \mathbf{x}_l)\,\delta(t - n\Delta t) \tag{1}$$

The electric and magnetic fields, $\mathbf{E}$ and $\mathbf{B}$, are defined on a spatial grid $\{\mathbf{x}_l\}$ and interpolated to the particle position $\mathbf{x}$ by means of interpolation functions $W$, typically products of splines. $W$ need not be, and often is not, the same for each field component. $\mathbf{v}$ is the particle velocity, and $\gamma = 1/\sqrt{1-v^2}$; $n$ is the time step index.

Eq. (1) can, of course, be integrated immediately to yield the standard leap-frog recurrence relation for $\mathbf{x}$ at integer times and $\mathbf{v}$ at half-integer times. However, as written, Eq. (1) emphasizes the similarity between real particles and numerical particles, and also leads naturally to the numerical dispersion relation illustrated in the next section. Currents, needed to advance the fields, can be expressed as grid quantities,

$$\mathbf{J}_l^{n+\frac{1}{2}} = \sum_p \mathbf{v}\, V(\mathbf{x} - \mathbf{x}_l)\,|_{t=(n+\frac{1}{2})\Delta t} \tag{2}$$

summed over particles. $V$ is another interpolation function. Eq. (2) does not conserve charge without correction, which involves solving Poisson's equation on the grid.[1] The modestly more complicated current algorithm developed by Esirkepov avoids this issue.[2]

Typically, fields are defined on Yee's FDTD mesh,[3] shown in Fig. 1. Magnetic fields, offset one-half time-step, are averaged in time to obtain values at integer times for Eq. (1). The difference equations are lengthy but straightforward. For instance, in 1D the transverse field equations take the form,

$$E_l^{n+1} = E_l^n - (B_{l+\frac{1}{2}}^{n+\frac{1}{2}} - B_{l-\frac{1}{2}}^{n+\frac{1}{2}})\frac{\Delta t}{\Delta x} - J_l^{n+\frac{1}{2}}\Delta t \tag{3}$$

$$B_{l+\frac{1}{2}}^{n+\frac{1}{2}} = B_{l+\frac{1}{2}}^{n-\frac{1}{2}} - (E_{l+1}^n - E_l^n)\frac{\Delta t}{\Delta x} \tag{4}$$

Staggered in space and time, the Yee field algorithm is second-order accurate in both. More complex expressions for Eq. (4) can reduce numerical dispersion.[4, 5, 6, 7]

Alternatively, fields can be evaluated as a set of modes in Fourier space, an algorithm known as PSTD.[8] Difference equations for the mode amplitudes are given by

$$\mathbf{E}^{n+1} = \mathbf{E}^n - i\mathbf{k} \times \mathbf{B}^{n+\frac{1}{2}}\Delta t - \mathbf{J}_k^{n+\frac{1}{2}}\Delta t \tag{5}$$

$$\mathbf{B}^{n+\frac{1}{2}} = \mathbf{B}^{n-\frac{1}{2}} + i\mathbf{k} \times \mathbf{E}^n\Delta t \tag{6}$$

Eq. (1) and (2) require fields and produce currents in real, not Fourier, space. Consequently, Fourier transforms of both fields and currents are required at each time step, which can be expensive. The fields and currents in real space typically are located at mesh corners, with $\mathbf{B}$ and $\mathbf{J}$ offset a half time step from $\mathbf{E}$. Fields are exact in space (for the Fourier modes computed) and second-order accurate in time.

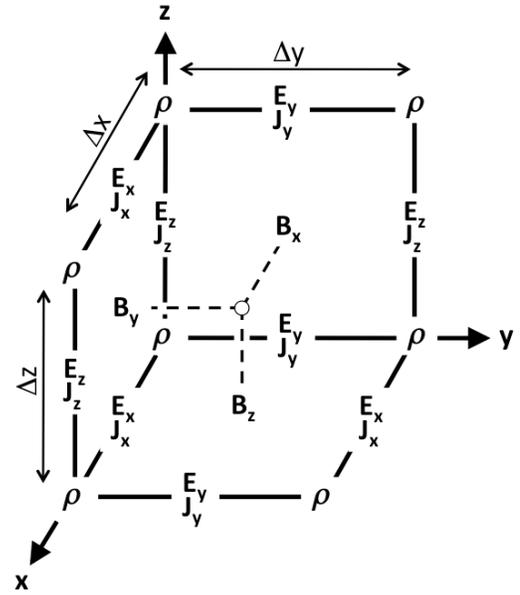

**Figure 1.** Yee's staggered mesh. Magnetic fields and currents are offset one-half time step.

A variant on PSTD, the Pseudo-Spectral Analytical Time Domain (PSATD) algorithm, was developed by Haber, et. al. some four decades ago.[9, 10] It can be derived by analytically integrating Maxwell's equations over a time step with the current held constant. Thus, it is exact for constant current. The resulting difference equations are

$$\mathbf{E}^{n+1} = \mathbf{E}^n - 2iS_h\mathbf{k}\times\frac{\mathbf{B}^{n+\frac{1}{2}}}{k} - \frac{2S_h C_h \mathbf{J}^{n+\frac{1}{2}}}{k} + \frac{2S_h C_h}{k}\mathbf{k}\mathbf{k}\cdot\frac{\mathbf{J}^{n+\frac{1}{2}}}{k^2} - \mathbf{k}\mathbf{k}\cdot\frac{\mathbf{J}^{n+\frac{1}{2}}\Delta t}{k^2} \tag{7}$$

$$\mathbf{B}^{n+\frac{1}{2}} = \mathbf{B}^{n-\frac{1}{2}} + 2iS_h\mathbf{k}\times\mathbf{E}^n\Delta t \tag{8}$$

with $S_h = \sin\frac{k\Delta t}{2}$, $C_h = \cos\frac{k\Delta t}{2}$, and $\boldsymbol{B}^n = (\boldsymbol{B}^{n+\frac{1}{2}} + \boldsymbol{B}^{n-\frac{1}{2}})/2C_h$.[11] It has many desirable properties but, like PSTD, requires Fourier transforms of the fields and currents at every time step.

Advancing the fields multiple times per particle time step (i.e., $\Delta t_{field} = \Delta t_{part}/N$) sometimes is advantageous. Sub-cycling the fields is numerically straightforward, of course. Sub-cycling can be treated analytically by casting the $N = 1$ difference equations for any algorithm into the form, $F^{n+1} = M:F^n + S^{n+\frac{1}{2}}$, where $F$ is a vector of fields, $S$ a vector involving source terms, and $M$ a matrix of difference terms. The corresponding expression for any $N$ easily is shown to be $F^{n+1} = M^N:F^n + \sum_{i=0}^{N-1} M^i:S$. The sum can be performed explicitly to yield the convenient

$$F^{n+1} = M^N:F^n + (M - I)^{-1}:(M^N - I):S \tag{9}$$

This analysis has been carried out for PSTD,[12] although the result is too lengthy to be presented here. The vacuum dispersion relation is, however, both short and interesting.

$$\sin^2\left(\frac{\omega \Delta t}{2}\right) - \sin^2\left(\frac{\theta \Delta t}{2}\right) = 0 \tag{10}$$

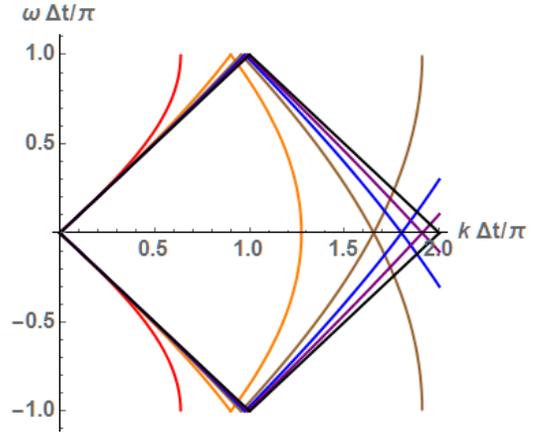

with $\sin\left(\frac{\theta \Delta t}{2N}\right) = \frac{k \Delta t}{2N}$. (From here onward, $\Delta t$ represents the particle time step.) Eq. (10) solutions for $N = 1, 2, 3, 4, 6, \infty$ are shown in Fig. 2. Of the many solution branches, only those observed by the particles, $|\omega| < \pi/\Delta t$, are plotted. The red $N = 1$ curve corresponds to PSTD, and the black $N = \infty$ curve to PSATD.[10] Indeed, the complete PSATD dispersion relation can be obtained from that of PSTD in the $N \to \infty$ limit.[12]

Eq. (10) has no real solutions for $\omega(k)$ when $\frac{k \Delta t}{2N} > 1$. More generally, electromagnetic numerical dispersion relations take the form

$$\sin\frac{\omega\Delta t}{2N} = f(\Delta x_i)\frac{\Delta t}{2N} \tag{11}$$

**Figure 2.** EM vacuum dispersion for $N = 1 - 4, 6, \infty$.

with the function $f$ of the cell dimensions depending on the details of the field solver. Because the left side of Eq. (11) cannot exceed unity for real $\omega$, $\Delta t$ is constrained to be less than $2N/f_{max}$, with $f_{max}$ the maximum value of $f$ on the grid. ($f$ almost always assumes its maximum value on the grid boundary.) Samples of this constraint, known as the Courant limit,[1] are for $\Delta x_i$ equal and grid dimension $d$,

- PSTD              $\Delta t_{max}/\Delta x = 2N/\pi\sqrt{d}$       {0.63, 0.45, 0.36}N for {d=1, 2, 3}
- FDTD (Yee)        $\Delta t_{max}/\Delta x = N/\sqrt{d}$           {1.00, 0.70, 0.57}N for {d=1, 2, 3}
- FDTD (C-K)        $\Delta t_{max}/\Delta x = N$                    {1.00, 1.00, 1.00}N for {d=1, 2, 3}
- PSATD             no limit

However, as apparent from Fig. 2, electromagnetic dispersion as seen by the particles becomes distorted for $\Delta t/\Delta x > 1/\sqrt{d}$ (more or less). It is prudent to introduce a degree of digital filtering for $\Delta t/\Delta x$ significantly larger than this.

The various numerical approximations inherent in PIC codes introduce a degree of inaccuracy. For instance, the much smaller number of particles per cell in a simulation as compared to a real plasma can introduce significant numerical noise. Moreover, simulations are only second order accurate in time and, for FDTD at least, in space too. This translates to errors of order $(\omega\Delta t)^2$ and $(k\Delta x)^2$, with $\omega$ and $k$ the frequencies and wavenumbers of physical interest.

Particles support more wavenumbers than fields do, because particles are distributed across the computational grid, whereas fields are evaluated at discrete points on the grid. This leads to numerical coupling among aliases, $k_P = (k_F + 2\pi m/\Delta x)$ for integer $m$. A second mismatch between particles and fields is this: The particle pusher is Lagrangian, and the field solver is Eulerian. This, too, can lead to spurious modes and aliases. Both drive numerical instabilities.

# BRIEF INTRODUCTION TO PIC NUMERICAL STABILITY ANALYSIS

Derivation of the linear dispersion relation for a simulation plasma parallels the derivation for a real plasma, although the calculation and resulting expression usually are more complicated.[13] First, the Vlasov equation based on Eq. (1) is linearized to obtain

$$\frac{\partial f}{\partial t} + \boldsymbol{v} \cdot \frac{\partial f}{\partial \boldsymbol{x}} = -\frac{\partial f^0}{\partial \boldsymbol{p}} \cdot \sum_{n,l} (\boldsymbol{E}_l^n + \boldsymbol{v} \times \boldsymbol{B}_l^n) W(\boldsymbol{x} - \boldsymbol{x}_l) \, \delta(t - n\Delta t) \tag{12}$$

Then, the linearized Vlasov equation is Fourier transformed in space and time,

$$f = i \frac{\partial f^0}{\partial \boldsymbol{p}} \cdot \frac{(\boldsymbol{E}_k^\omega + \boldsymbol{v} \times \boldsymbol{B}_k^\omega) W_{-k'}}{\omega' - \boldsymbol{k}' \cdot \boldsymbol{v}} \tag{13}$$

where $\omega' = \omega + n\frac{2\pi}{\Delta t}$ and $\boldsymbol{k}' = \boldsymbol{k} + \boldsymbol{m}\frac{2\pi}{\Delta x}$ are particle aliases of the field modes, as explained at the end of the previous section.

Similarly, Eq. (2) is Fourier transformed to obtain

$$\boldsymbol{J}_k^\omega = i \int d^3v \, \boldsymbol{v} \, \frac{\partial f^0}{\partial \boldsymbol{p}} \cdot \sum_{n,m} (-1)^n \frac{V_{k'} (\boldsymbol{E}_k^\omega + \boldsymbol{v} \times \boldsymbol{B}_k^\omega) W_{-k'}}{\omega' - \boldsymbol{k}' \cdot \boldsymbol{v}} \tag{14}$$

The factor $(-1)^n$ occurs, because currents are defined at half-integer time steps.[13] The charge density, defined at integer time steps, does not contain this factor.

$$\rho_k^\omega = i \int d^3v \, \frac{\partial f^0}{\partial \boldsymbol{p}} \cdot \sum_{n,m} \frac{V_{k'} (\boldsymbol{E}_k^\omega + \boldsymbol{v} \times \boldsymbol{B}_k^\omega) W_{-k'}}{\omega' - \boldsymbol{k}' \cdot \boldsymbol{v}} \tag{15}$$

As mentioned earlier, $\boldsymbol{J}$ as defined in Eq (2) does not conserve charge and, therefore, must be corrected in order that Gauss' law remain satisfied. The process, which requires $\rho$, is described in Ref. 1, 11, and elsewhere. The more complicated Esirkepov algorithm[2] is to be preferred, because it obviates the need for computing $\rho$, and correcting the current at each time step. Also, numerical instability growth rates typically are modestly smaller for the Esirkepov algorithm.[14] (See Eq. (7) and associated text in Ref. 14 for the term replacing "$\boldsymbol{v}$" in Eq. (14) of this article.)

Finally, the current is substituted into the Fourier-transformed finite difference field equations described in the previous section to obtain a dispersion matrix. Its determinant is the desired dispersion relation.

A cursory examination of Eq. (14) reveals several potential sources of numerical instability. The denominator contains an infinite family of nonphysical resonances, $\omega' - \boldsymbol{k}' \cdot \boldsymbol{v}$ ($\boldsymbol{m} \neq 0$) that can couple unstably with physical resonances and, depending on the form of $f^0$, with one another. Moreover, the coefficient of the resonance, $V_{k'} W_{-k'}$, may have the wrong sign, triggering "Landau growth" instead of damping (for instance), again depending on the form of $f^0$. Although perhaps not immediately obvious from Eq. (14), the numerator, $\boldsymbol{E}_k^\omega + \boldsymbol{v} \times \boldsymbol{B}_k^\omega$, contains a term qualitatively of the form, $\frac{\sin(\omega \Delta t/2)}{\Delta t/2} - \frac{\sin(k\Delta x/2)}{\Delta x/2} \boldsymbol{v}$, which should cancel with the $n = \boldsymbol{m} = 0$ resonance in the denominator, $\omega - \boldsymbol{k} \cdot \boldsymbol{v}$, but does so only in the limit of vanishing time step and cell size. This spurious, predominantly transverse mode (and its aliases) should not be confused with the physical, predominantly longitudinal mode (and its aliases). For a relativistic beam, the former scales as $\omega_p$, while the latter scales as $\omega_p/\gamma$. (Here, the beam plasma frequency, $\omega_p$, is defined to contain a factor of $\gamma^{-\frac{1}{2}}$.) It, too, can couple unstably with physical resonances. Finally, violating the Courant limit causes strong instability. Note that finite plasma density makes the Courant limit slightly more restrictive and sometimes even causes weak, narrow band instabilities near $\omega = \pm 2\pi/\Delta t$; see Fig. 2.[1, 11]

Several methods are employed to minimize PIC numerical instabilities. Because numerical instabilities usually occur at wavelengths of two to four cells, they can be suppressed by digitally filtering those wavelengths. This is easy for PSTD: Just set the high-*k* current modes to zero. More or less equivalently, set to zero the high-*k* field modes when interpolating them to the particles. The corresponding approach for FDTD, repeatedly smoothing short wavelength current and field fluctuation by means of a (¼,½,¼) stencil, does not produce a sharp cutoff but still is useful.[15] Employing higher order interpolation functions also is helpful. Except at small $k'$, $V_{k'} W_{-k'}$ varies as $k'^{-4}$ for linear interpolation and as $k'^{-8}$ for cubic interpolation. Thus, cubic interpolation is quite effective at suppressing higher order aliases, although much less so for $m = 0, -1$. Sometimes, simply choosing simulation parameters wisely works. For instance, FDTD beam simulations are substantially more stable at certain "magic time steps".[14]

In many instances, modifying the algorithm itself is warranted. Implicit algorithms usually introduce a degree of damping, especially at short wavelengths, that reduces numerical instability growth rates.[15] Alternative, less drastic, modifications are discussed in the next section.

## SUPPRESSING THE NUMERICAL CHERENKOV INSTABILITY

The most serious numerical instability in multidimensional PIC simulations of relativistic beams (in high-current accelerators, astrophysical shocks, etc.) is the numerical Cherenkov instability, so named because it results from the interaction of spurious beam modes with numerically distorted electromagnetic modes.[13, 16] Its peak growth rate can be a large fraction of $\left(\omega_p^2 k_\perp^2 \Delta z\right)^{1/3}$, where $\Delta z$ is the cell dimension in the direction of beam propagation. For highly relativistic beams, its dispersion relation is

$$C_0 + \omega_p^2 \sum_m C_1 \csc\left[(\omega - k_z' v)\frac{\Delta t}{2}\right] + \omega_p^2 \sum_m C_{2x} \csc^2\left[(\omega - k_z' v)\frac{\Delta t}{2}\right] = 0 \qquad (16)$$

and $C_{2x}$ are functions of $\omega$, $\mathbf{k}$, and the simulation parameters, and $C_0$ is the vacuum numerical dispersion relation. $C_1$ Details depend on the algorithm chosen; e.g., FDTD,[14] PSTD,[12] or PSATD.[11]

To understand the instability, it is helpful to visualize the locations of the modes involved, and where they intersect. The left chart in Fig. 3 shows normal modes for 2D PSATD with $\frac{v\Delta t}{\Delta z} = 1.2$, $v \approx 1$, $\Delta z = \Delta x = 0.3868$. This particular slice through phase space is for $k_x = 1$, or about one-eighth the maximum value of $k_x$ on the grid. Note that the electromagnetic modes (dashed lines) turn over at large $k$, as described in the previous section. For somewhat smaller $\frac{v\Delta t}{\Delta z}$, the electromagnetic modes no longer intersect the $m = 0$ beam mode.[11] Nonetheless, they still intersect aliases in various places. The locations in $k$-space of these intersections for $m = -1, 0, +1$ are shown as black curves in the chart on the right. Superimposed are instability growth rates for $\omega_p = 1$ and linear interpolation.

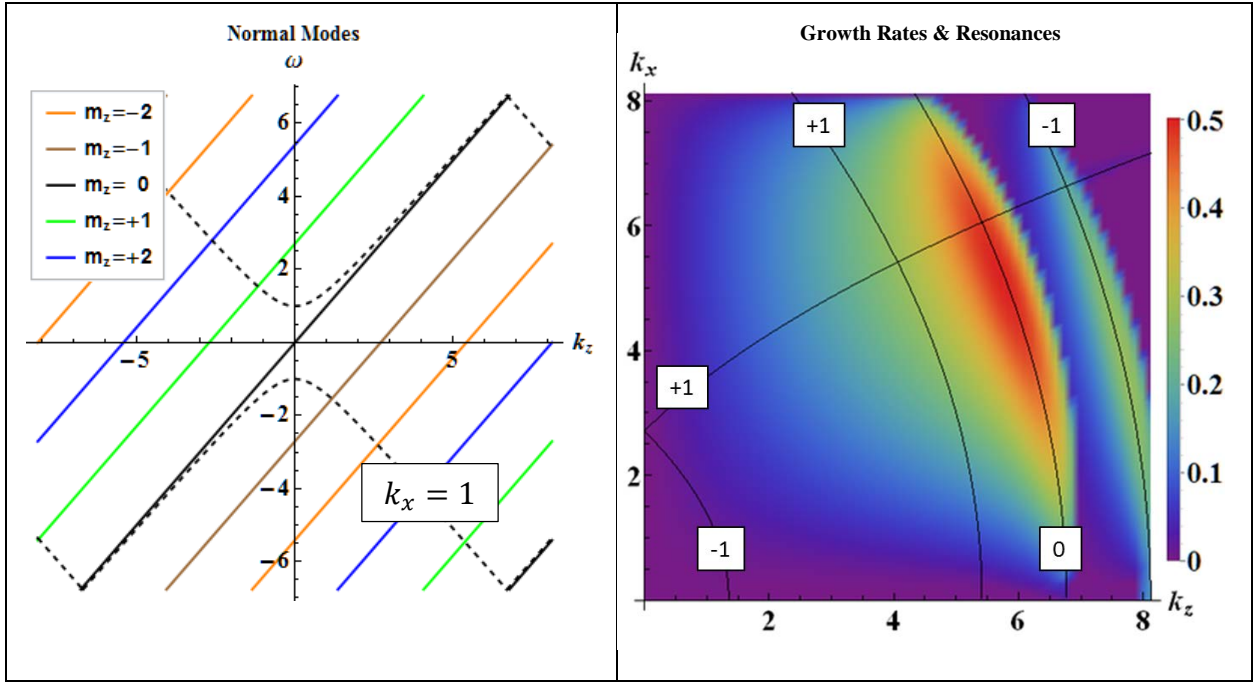

**Figure 3.** Normal Modes (left) and Growth Rates & Resonances (right) for PSATD with $\frac{v\Delta t}{\Delta z} = 1.2$, $v \approx 1$.

Normal mode diagrams for PSTD and FDTD are qualitatively similar but differ from the left side of Fig. 3 as follows. PSTD electromagnetic modes travel faster than the speed of light (unity for the normalization used in this paper) and bend upward relative to the PSATD curves. This also is apparent in Fig. 2. FDTD electromagnetic modes, on the other hand, travel slower than the speed of light at large $k$,[14] bending downward relative to the PSATD curves. Hence, FDTD electromagnetic modes almost always intersect the $m = 0$ beam mode. Lehe's variant of FDTD is an exception.[7] Nonetheless, in all these algorithms electromagnetic modes interact with spurious beam modes.

The right side of Fig. 3 displays two noticeable features, multiple resonant numerical instabilities in narrow bands at large $k$, and a slower growing nonresonant instability in a wide band over smaller $k$. As stated earlier, higher order resonances can be weakened substantially by using cubic interpolation. (Employing still higher order interpolation functions offers little advantage, however.) Suppressing the $m = 0, -1$ resonant instabilities typically requires digital filtering of short wavelengths, and there are several ways to accomplish this.[11, 14, 17, 18, 19, 20, 21] The nonresonant instability, on the other hand, extends into regions of $k$-space where physical phenomena of interest may occur, so digital filtering is not practical. Instead, the nonresonant instability can be suppressed by making minor corrections to the fields as interpolated to the particles, or sometimes by making minor corrections to the currents.

Interpolated-field corrections that minimize $C_{2x}$ in Eq. (16) for $\omega \approx \boldsymbol{k} \cdot v$ are considered below. Essentially, the corrections cancel some numerical errors introduced by finite cell size and time step and, therefore, would not be expected to distort physical phenomena. (Recall that $C_{2x}$ vanishes in the limit of small cell size and time step.) Correction factors differ from unity by of order $k^2$ for small $k$. Because the interpolated-field correction factors, defined as $\Psi_E$ and $\Psi_B$, enter linearly into $C_{2x}$, only their ratio is determined.

Several FDTD interpolation schemes exist, and each has its own ratio, $\Psi_E/\Psi_B$.[18] For instance, the Galerkin "energy-conserving" scheme has as a ratio,

$$\Psi_E/\Psi_B = \sin\left(k_z \frac{\Delta t}{2}\right) \cos\left(k_z \frac{\Delta t}{2}\right) \cot\left(k_z \frac{\Delta z}{2}\right) \frac{\Delta z}{\Delta t} \quad (17)$$

This and corresponding expressions for other interpolation schemes can be satisfied exactly in $k$-space, but the Fourier transforms at every time step incur unnecessary cost. Instead, Eq. (17) can be approximated to an accuracy of about $10^{-6}$ by the ratio of two fourth-order polynomials in $\sin^2\left(k_z \frac{\Delta z}{2}\right)$. $\Psi_E$ and $\Psi_B$ then can be set to the numerator and denominator, respectively; $\sin^2\left(k_z \frac{\Delta z}{2}\right)$ is computed by applying the (-¼, ½, -¼) stencil to a copy of the fields before interpolating them to the particles.

Sample results are provided in Fig. 4, based on cubic interpolation, Cole-Karkkainen FDTD field solver, and two-pass bilinear filtering.[18] Other parameters are as in Fig. 3. WARP[22] simulation results are provided for comparison. Instability growth rates are less than 0.01 $\omega_p$ throughout most of the plot, more than an order of magnitude smaller than without the field correction procedure. WARP LPA simulations using field corrections are devoid of visible numerical issues and produce physical results as expected.[18]

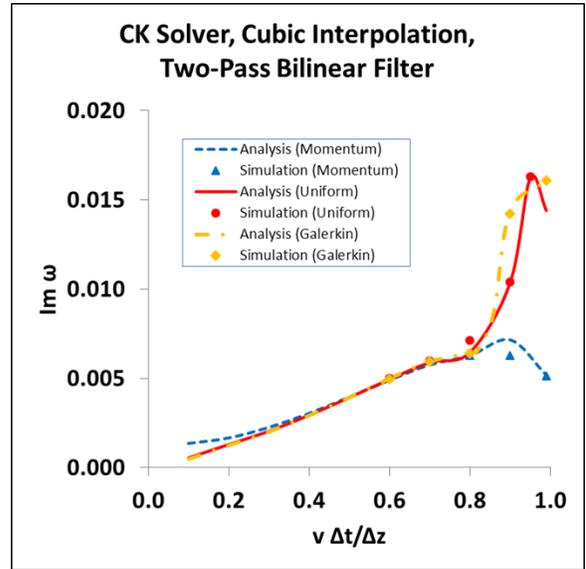

**Figure 4**. Maximum growth rates from FDTD numerical dispersion relation, compared with simulation results.

The Sub-Cycled PSTD $\Psi_E/\Psi_B$ ratio is somewhat more complicated.[12]

$$\Psi_E/\Psi_B = \frac{\sin\left(k_z v \frac{\Delta t}{2}\right) \cos\left(k_z v \frac{\Delta t}{2}\right) \sin^2\left(\theta \frac{\Delta t}{2}\right)}{\left(k_z v \frac{\Delta t}{2}\right)\left[\sin^2\left(\theta \frac{\Delta t}{2}\right) - \sin^2\left(k_z v \frac{\Delta t}{2}\right)\left(1 - \frac{\sin(\theta \Delta t)}{N} \csc\left(\frac{\theta \Delta t}{N}\right)\right)\right]} \quad (18)$$

The right side of Eq. (18) can be split arbitrarily between $\Psi_E$ and $\Psi_B$, provided that each differs from unity by of order $k^2$ or less for small $k$. In the discussion that follows, $\Psi_B$ is set to unity. Setting $\Psi_E$ to unity instead typically gives very similar results away from the singularity in Eq. (18). It is advisable to use a sharp cutoff digital filter to avoid the effects of the singularity as well as to eliminate most of the residual $m = -1$ instability.[17] Specifically, $\Psi_E$ and $\Psi_B$ are set to zero for $k > \alpha \min\left[\frac{\pi}{\Delta z}, \frac{\pi}{v \Delta t}\right]$, with $\alpha < 1$. The left chart in Fig. 5 depicts maximum growth rates for various $N$ as functions of $\Delta t$ for $\Psi_E$ and $\Psi_B$ as just defined, $\alpha = 0.6$, and a single pass bilinear filter applied the fields and currents. (Similar results are obtained for α as large as 0.75.)

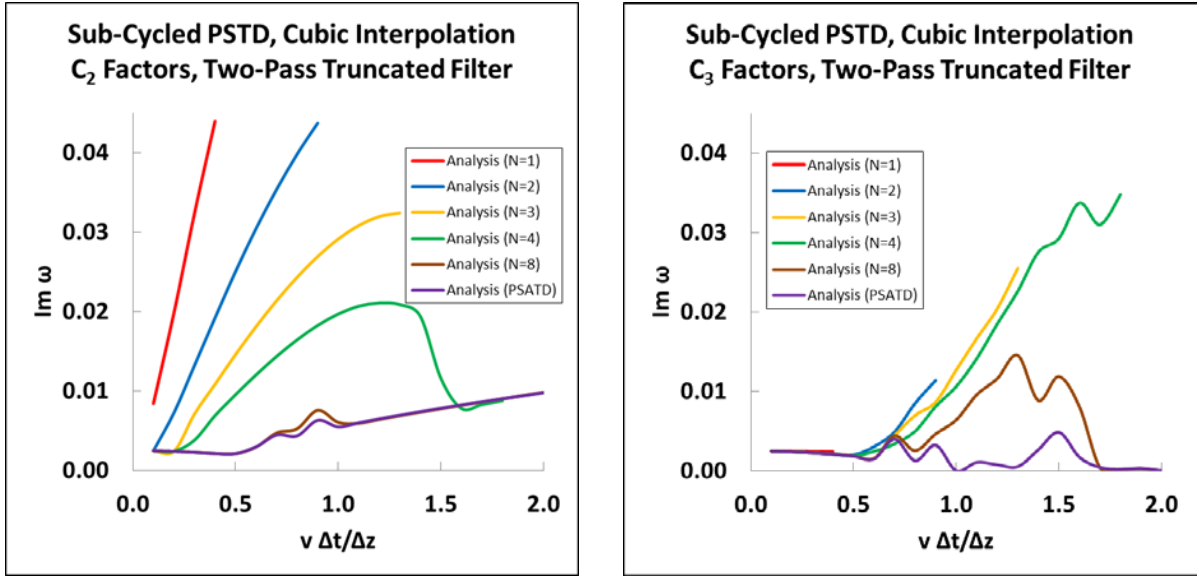

**Figure 5.** Maximum growth rates from PSTD numerical dispersion relation for $C_{2x}$ and $C_{3x}$ correction factors.

Growth rates are larger in Fig. 5 than in Fig. 4 primarily due to a weak, narrow band instability,[11, 20] which is associated primarily with the $C_1$ term in Eq. (16). It typically occurs in PSTD at $k_z$ about half the Nyqyist wavenumber and small $k_x$ (Ref. 17 mistakenly associated it with $C_{2x}$.) Yu, *et. al.* suggest that this residual instability be eliminated by a combination of small $\Delta t$ and digital filtering.[21]

An alternative to Eq. (18), here designated $C_{3x}$, is

$$\Psi_E = \left(k_z \frac{\Delta t}{2}\right) \cot\left(k_z \frac{\Delta t}{2}\right), \quad \Psi_B = \left(k \frac{\Delta t}{2}\right) \cot\left(\theta \frac{\Delta t}{2}\right) \sec\left(\frac{\theta \Delta t}{2N}\right) \quad (19)$$

The PSATD limit was used successfully in Ref. 18, where it is referred to as method "b2". The right chart in Fig. 5 shows that Eq. (19) gives slightly better results than does Eq. (18), at least for the parameters chosen.

As proposed by Vay, *et. al.*, PSTD can be generalized further by replacing the components of $\mathbf{k}$ in the field and Esirkepov current algorithms by the Fourier transforms of various order finite difference approximations to spatial derivatives on the grid.[24] (In this context the components of $\mathbf{k}$ itself can be viewed as infinite order approximations.) Dubbed the Pseudo Spectral Arbitrary Order Time Domain (PSAOTD) algorithm, it is intended to provide more localized particle representations than does standard PSTD.[25] This research is in a preliminary state.

Figure 6 provides PSTD peak growth rates with $C_{2x}$ correction factors for $N = 1, 2, 4, 8$ and orders 2, 4, 6, 8, 10, and infinity. $C_{2x}$ remains as defined in Eq. (18) but with the definition of $\theta$ modified to

$$\sin\left(\frac{\theta \Delta t}{2N}\right) = \frac{\sqrt{[k_x]^2 + [k_y]^2 + [k_z]^2} \, \Delta t}{2N} \quad (20)$$

with $[k_i]$ the appropriate finite difference approximation to $k_i$. Digital filtering is as in Fig. 5.

Not surprisingly, the dispersion relation for $N = 1$, order 2 PSTD is essentially the same as that for uniform interpolation FDTD. Hence, peak growth rates for these two algorithms, shown in Fig. 4 and the upper left chart of Fig. 6, have similar values over the range $v\frac{\Delta t}{\Delta z} < \frac{1}{\sqrt{2}}$. (Incorporating the equivalent of the Cole-Karkkainen field solver into PSAOTD to permit comparing the two figures throughout the range $v\frac{\Delta t}{\Delta z} < 1$ would not be difficult. Agreement might not be as good at large $\Delta t$, however, because the rational approximation leading to Fig. 4 is less accurate there.)

As in Fig. 5, here too the larger than desired residual growth rates are due to the weak, narrow band PSTD instability. This instability is weaker for large $N$. Evidently, it also does not occur for order 2 at any $N$. In contrast to the results of the right side of Fig. 5, $C_{3x}$ field correction does not work well except in narrow windows unless $N$ or order is large.

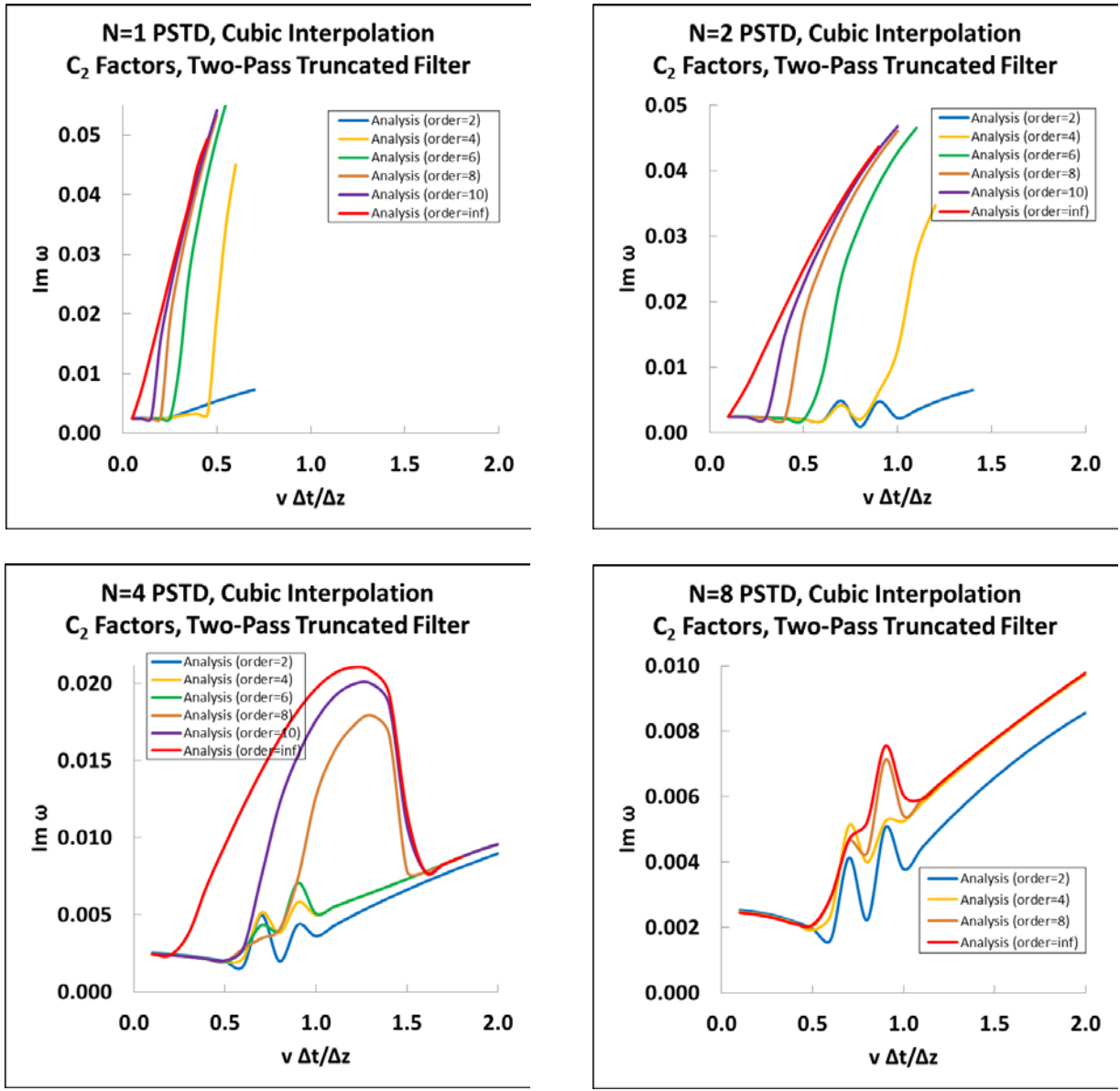

**Figure 6.** Maximum growth rates from PSTD numerical dispersion relation with $N = 1, 2, 4, 8$ and various orders for $C_{2x}$ correction factors.

Some of the software used to produce the results in this section, along with a number of related presentations, is available at http://hifweb.lbl.gov/public/BLAST/Godfrey/, and more is to be added.

## ACKNOWLEDGMENTS

Many of the concepts discussed in this paper first were proposed by Jean-Luc Vay. He also performed WARP simulations validating much of the analytical modeling here. Although not discussed in the paper, the simulations were vital to progress. This work was supported in part by US-DOE Contract DE-AC02-05CH11231.